\documentclass[11pt]{JHEP3}

\preprint{hep-ph/0212263\\SISSA 95/2002/EP}
\keywords{Renormalization Group, Beyond Standard Model, Neutrino Physics}

\usepackage{amsmath}
\usepackage{epsfig,bbold}

\newcommand{\U}{\mathop{\rm {}U}\nolimits}
\newcommand{\diag}{\mathop{\rm diag}\nolimits}

\title{Radiative corrections to neutrino mass matrix
in the Standard Model and beyond}

\author{Michele Frigerio\\
INFN, Section of Trieste and International  School
for Advanced Studies (SISSA)\\
Via Beirut 4, I-34013 Trieste, Italy\\
E-mail: \email{frigerio@he.sissa.it}}

\author{Alexei Yu.\ Smirnov\\
The Abdus Salam International  Center for Theoretical Physics (ICTP)\\ 
I-34100 Trieste, Italy, and\\
Institute for  Nuclear Research,  Russian Academy of Sciences\\ 
Moscow, Russia\\
E-mail: \email{smirnov@ictp.trieste.it}}

\abstract{We study the effect of radiative corrections on the
structure of neutrino mass matrix.  We analyze the renormalization of
the matrix from the electroweak scale $m_Z$ to the scale $m_0$ at
which the effective operator that gives masses to neutrinos is
generated.  Apart from Standard Model and MSSM, non-standard
extensions of SM are considered at a scale $m_X$ intermediate between
$m_Z$ and $m_0$.  We find that the dominant structure of the neutrino
mass matrix does not change. SM and MSSM corrections produce small
(few percents) independent renormalization of each matrix element.
Non-standard (flavor changing) corrections can modify strongly small
(sub-dominant) matrix elements, which are important for the low energy
phenomenology.  In particular, we show that all sub-dominant elements
can have purely radiative origin, being zero at $m_0$.  The set of
non-zero elements at $m_0$ can be formed by (i) diagonal elements
(unit matrix); (ii) $M_{ee}$ and $M_{\mu\tau}$; (iii)$M_{ee}$ and
$\mu\tau$-block elements; (iv) $\mu\tau$-block elements.  In the case
of unit matrix, both atmospheric and solar mixing angles and mass
squared differences are generated radiatively.}

\begin{document}

\section{Introduction}

Experiments on neutrino oscillations~\cite{exp}, neutrinoless $2\beta$
decay~\cite{2beta} and tritium $\beta$ decay~\cite{tritium} give
information on neutrino masses and mixing angles. In principle, also
CP violating phases can be measured. Using this information, it is
possible to reconstruct (at least partially) the Majorana mass matrix
of neutrinos at the electroweak scale $m_Z$.

The origin of the neutrino mass matrix is, most probably, in new
physics above a very high scale $m_0$, where the lepton number is
violated.  To find the structure of the matrix at $m_0$, it is
necessary to take into account the renormalization effects between
$m_Z$ and $m_0$.

The Renormalization Group Equation (RGE) for the neutrino mass matrix
has been extensively studied both in the Standard Model (SM) and
Minimal Supersymmetric Standard Model (MSSM)~\cite{MSSM}--\cite{many}.
The RGE has been also considered, in the context of the see-saw
mechanism, for non-degenerate heavy right-handed
neutrinos~\cite{munich}.  Low energy threshold corrections have been
computed; their effect can be important in the MSSM case~\cite{thre}.
A non-standard source of neutrino masses (operators in the K\"ahler
potential) has been considered in~\cite{CEN} and the corresponding
radiative corrections have been studied.  The goal of these analyses
was to understand how radiative corrections could affect the mass
squared differences $\Delta m^2_{ij}$ and the mixing angles
$\theta_{ij}$ observed in neutrino oscillation experiments. A general
conclusion is that the effect of running is small for hierarchical
mass spectra, while, in the case of quasi-degenerate mass spectrum,
the observables can be strongly modified by radiative corrections.  In
particular, in the degenerate case $\Delta m^2_{\rm sol}$ can have purely
radiative origin~\cite{petsmi}.

Mass squared differences and mixing angles are the outcome of the
diagonalization of the mass matrix and it is this matrix 
the object more closely related to the underlying theory. 
In contrast with previous studies, we will  analyze in detail
renormalization effects on the mass matrix structure.

In section~\ref{ope} we discuss the generation of neutrino mass matrix
at the scale $m_0$.  We study first the radiative corrections in SM
and MSSM (section~\ref{sms}).  Then, we consider other extensions of
the SM at some intermediate scale $m_X$, with $m_Z < m_X < m_0$
(section~\ref{newp}).  We identify the features of the mass matrix
which can be explained by radiative corrections.  Finally
(section~\ref{appli}), we consider four specific matrix structures at
$m_0$: (i) the matrix proportional to the unit; (ii) the matrix with
only $ee$ and $\mu\tau$ elements different from zero; (iii) the matrix
with non-zero $M_{ee}$ and $\mu\tau$-block elements; (iv) the matrix
with non-zero $\mu\tau$-block.  We will study the predictions for the
low energy parameters in the case of standard and non-standard
radiative corrections.

\section{Generation of the neutrino mass matrix and scale $m_0$}\label{ope}

With the SM fields, one can construct a unique Lorentz and gauge
invariant effective five dimensional operator that gives Majorana
masses to neutrinos~\cite{wein}:
\begin{equation}
\frac{C_{\alpha\beta}}{m_0}(\overline{L^c}_\alpha ~i\sigma_2~\phi)
(\phi i\sigma_2 L_\beta)+\hbox{h.c.}
\label{five}
\end{equation} 
Here $C_{\alpha\beta}$ are dimensionless couplings, $m_0$ is the mass
scale at which this effective interaction is generated, $\alpha,\beta$
are flavor indexes, $\phi$ and $L$ are the SM Higgs and lepton
doublets.  An analogue operator appears in the case of MSSM (without
R-parity violation).  After the electroweak symmetry breaking, the
operator~(\ref{five}) generates the neutrino mass matrix:
$$
M_{\alpha\beta}=\dfrac{2 C_{\alpha\beta}\langle \phi^0 \rangle ^2}{m_0}\,.
$$

Different possible mechanisms can lead to the effective
operator~(\ref{five}).  One possibility is the exchange of heavy
right-handed neutrinos (type-I see-saw~\cite{see1}).  In this case
$m_0$ should be identified with the mass of the lightest right-handed
neutrino $m_R$ and one gets
\begin{equation}
C_{\alpha\beta}= - \dfrac{m_0}{2}(Y_\nu M_R^{-1} Y_\nu^T)_{\alpha\beta}\,,
\label{seeco}
\end{equation}
where $Y_\nu$ is the matrix of neutrino Dirac Yukawa couplings and
$M_R$ is the Majorana mass matrix of right-handed neutrinos, both
evaluated at the scale $m_0=m_R$.

Another possibility is to introduce a scalar isotriplet $\Delta$ with
hypercharge $1$ and renormalizable coupling to the SM lepton doublets:
\begin{equation}
Y_{\alpha\beta}\overline{L^c}_\alpha 
\sigma^i L_\beta \Delta_i + \hbox{h.c.}\,.
\label{four}
\end{equation} 
A heavy triplet ($m_\Delta \gg m_Z$) can get a small induced VEV
due to the coupling
$M_{\Delta\phi}\Delta^*\phi\phi + \hbox{h.c.}$ (type-II see-saw~\cite{see2}):
$$
\langle\Delta^0\rangle = \frac {M_{\Delta\phi} 
\langle \phi^0 \rangle ^2}{m_\Delta^2}\,.
$$
If the exchange of the triplet is the dominant contribution to the neutrino
mass matrix, $m_0$ should be identified with $m_\Delta$ and 
\begin{equation}
C_{\alpha\beta}= Y_{\alpha\beta}\frac{M_{\Delta\phi}}{ m_0}\,.
\label{trico}
\end{equation}

The following remarks are in order:
\begin{enumerate}

\item
The neutrino mass matrix can receive both contributions of the
form~(\ref{seeco}) and~(\ref{trico}).  In this case $m_0 = \min \{ m_R
, m_\Delta \}$.

\item
The running between $m_0 = \min \{ m_R , m_\Delta \}$ and $\tilde{m}_0
= \max \{ m_R , m_\Delta \}$ can be important for the neutrino mass
matrix structure, if $N_R$ and $\Delta$ contributions are comparable.

\item
There can be a hierarchy among the masses of right-handed neutrinos,
so that the operator~(\ref{five}) is formed in a large energy
interval.  Corrections in this interval can be
important~\cite{munich}.

\end{enumerate}
The relative size and the structure of the various ($N_R$, $\Delta$,
maybe some other) contributions to the neutrino mass matrix are
model-dependent and deserve a separate analysis. In this paper we will
study the running below $m_0$.

\section{Renormalization of the mass matrix in the SM (MSSM)}\label{sms}

Let us consider, first, the renormalization of the mass matrix when
the only particles with mass below $m_0$ are the SM (MSSM) particles.
The $\beta$-function of the operator~(\ref{five}), $\beta_M \equiv \mu
\frac{\mathrm{d}}{\mathrm{d}\mu}M$, can be written
as~\cite{MSSM,adklr}:
\begin{eqnarray}
16\pi^2\beta_M^{SM}&=&-\dfrac 32 \left[M (Y^\dag_l Y_l) + (Y^\dag_l Y_l)^T M
\right]+ K_{SM}M\,,
\nonumber\\
16\pi^2\beta_M^{MSSM}&=& \left[M (Y^\dag_l Y_l) + (Y^\dag_l Y_l)^T M
\right]+ K_{MSSM}M\,,
\label{beta}
\end{eqnarray}
where $Y_l$ is the $3\times 3$ matrix of charged lepton Yukawa
couplings and $K_{SM(MSSM)}$ is a real parameter describing flavor
universal radiative corrections.

The flavor non-universal corrections (terms in square brackets in
eq.~(\ref{beta})) come from two types of diagrams only
(figure~\ref{stand}), generated by charged lepton Yukawa couplings:
renormalization of the wavefunction of lepton doublets ($a$) and
vertex correction ($b$).  In the SM, the coefficient $-3/2$ in
eq.~(\ref{beta}) is the sum of a contribution $1/2$ from (a) and $-2$
from ($b$).  In the MSSM, only ($a$) contributes, due to SUSY
non-renormalization, and the coefficient is $1/2 \times 2 = 1$, where
the factor $2$ corresponds to the double number of particles with
respect to SM.

\FIGURE[t]{\epsfig{file=stand.eps,width=.65\textwidth}%
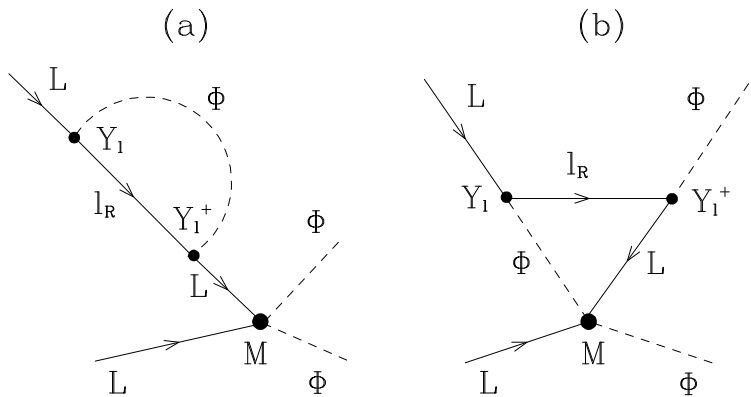
\caption{The flavor non-universal diagrams in the case of SM
(MSSM).\label{stand}}}

In flavor basis, the matrix $Y_l$ is
diagonal and it can be made real by reabsorbing the phases in the fields:
$$ 
Y^\dag_l Y_l = \diag(y_e^2,y_\mu^2,y_\tau^2) \,. 
$$
Furthermore, $y_{\alpha}$, taken real at a certain scale, will remain
real during the running~\cite{chapok}.  In this basis, the RGE
equation for $M_{\alpha\beta}$ can be easily integrated,
giving~\cite{ello}:
\begin{equation}
M_{\alpha\beta}(m_Z)=I_K\exp\left[-k\int_0^{t_Z} (y^2_{\alpha}(t)+
y^2_{\beta}(t)) dt \right] M_{\alpha\beta}(m_0)\,,
\label{mrun}
\end{equation}
where 
$$
t_Z\equiv \frac{\log(m_0/m_Z)}{16\pi^2}\,,\qquad
k=-\frac{3}{2}\quad{\rm (SM)}\,,\qquad k=1\quad{\rm (MSSM)}\,.
$$
The flavor universal corrections are contained in the prefactor
$$
I_K = \exp \left(-\int_0^{t_Z} K(t) dt \right).
$$ 
$I_K$ does not change the structure of the mass matrix and, moreover,
the overall renormalization effect is small: $I_K \approx 1$. We will
consider only 1-loop radiative corrections and therefore neglect the
evolution of $y_\alpha$ in eq.~(\ref{mrun}), taking the values of
charged lepton Yukawa couplings at the electroweak scale:
$y_\alpha(t)\approx y_\alpha(0)$.

Several important conclusions follow immediately from eq.~(\ref{mrun}): 
\begin{itemize}

\item In flavor basis, each element of the mass matrix evolves
independently from the values of other elements.  The value of the
element at $m_Z$ is proportional to its value at $m_0$:
\begin{equation}
M_{\alpha\beta}(m_Z)\propto M_{\alpha\beta}(m_0)\,.
\label{propto}
\end{equation} 
In particular, elements which are zero at some scale, 
remain zero at any scale.

\item The phases of $M_{\alpha\beta}$ do not evolve,
because the r.h.s.\ in eq.~(\ref{mrun}) is real. 
In contrast, the three physical CP violating phases depend on the moduli
of $M_{\alpha\beta}$ and therefore can be strongly renormalized.
However, if there is no CP violation at some scale ($M$ is real),
no CP violation can be induced by radiative corrections at any scale.

\item The corrections to matrix elements have opposite sign in SM and
MSSM, due to the different sign of $k$.  The same matrix at $m_Z$ will
develop different features at the scale $m_0$ depending on whether low
energy supersymmetry is present or not.

\end{itemize}
Notice that, in principle, SM and MSSM could be discriminated by the
ordering of mass eigenvalues: exactly degenerate neutrinos at the
scale $m_0$ could be split into a normal mass spectrum in SM and into
an inverted spectrum in MSSM or vice versa, depending on mixings and
phases. However, using only SM (MSSM) radiative corrections, it is
hard to reproduce data starting with exactly degenerate
neutrinos~\cite{dj}.  Also the scenario with only two degenerate
neutrinos at high energy has been recently studied~\cite{jr} and the
difference of predictions between SM and MSSM has been analyzed.

The largest flavor dependent contribution to the running in eq.~(\ref{mrun})
is due to the $\tau$ Yukawa coupling: 
\begin{equation} 
y_{\tau}(m_Z) = \left\{ \begin{array}{ll}
\dfrac{\sqrt{2}m_\tau}{v}\approx 10^{-2}  &  
{\rm (SM)}\\ \dfrac{\sqrt{2}m_\tau}{v\cos\beta} \approx
\tan\beta\cdot 10^{-2} &  {\rm (MSSM)}
\end{array} \right.\,,
\end{equation}
where $v\approx 246$\,GeV is the VEV of the SM Higgs and $\tan\beta$
is the ratio of VEV's of the two MSSM Higgs doublets.  The other
Yukawa couplings are much smaller, therefore the largest corrections
are for elements which have the $\tau$-flavor.  Neglecting $y_e$ and
$y_\mu$ corrections, eq.~(\ref{mrun}) gives:
\begin{eqnarray*}
M_{\tau\tau}(m_Z)&\approx& I_K M_{\tau\tau}(m_0)(1-2k\epsilon_\tau)\,,
\\
M_{e\tau}(m_Z)&\approx& I_K M_{e\tau}(m_0)(1-k\epsilon_\tau)\,,
\\
M_{\mu\tau}(m_Z)&\approx& I_K M_{\mu\tau}(m_0)(1-k\epsilon_\tau)\,,
\end{eqnarray*}
where, taking $m_0=10^{n}{\rm ~GeV}$,
\begin{equation}
\epsilon_\tau \equiv \frac{y_{\tau}^2(m_Z)}{16\pi^2} \log\frac{m_0}{m_Z} 
\left\{ \begin{array}{ll} \approx 1.5(n-2)10^{-6} &  \textrm{(SM)}
\\
\approx 3.5(n-2)10^{-3}\left(\dfrac{\tan\beta}{50}\right)^2 &  
\textrm{(MSSM)}
\end{array} \right.\,.
\label{epsi}
\end{equation}
The effect of running is apparently very small even
for $M_{\tau\tau}$.

We conclude that radiative corrections (in SM and MSSM) practically do
not change the structure of neutrino mass matrix up to the scale
$m_0$.  Zero elements remain zero and non-zero elements acquire very
small relative corrections, which are at most few percents.  Symmetry
properties of the matrix at $m_0$ are almost unchanged by running to
low energy, where the matrix structure can be reconstructed using
experimental input for mixing angles and mass eigenvalues, as well as
CP violating phases~\cite{MAN,MAI}. In other words, the structure of
the mass matrix is stable with respect to radiative corrections in SM
and MSSM, independently of type of mass ordering (normal or inverted),
level of degeneracy, values of mixing angles and CP violating phases.

\subsection{Radiative corrections and observables}

Even though the matrix structure is stable, observables, i.e.\ the
values of the mass differences and the form of the mixing matrix, can
be strongly affected by the radiative corrections.  There is a number
of studies~\cite{chapok}--\cite{many} in which the RGE's for mixing
angles and $\Delta m^2_{ij}$ have been analyzed and conditions for
strong renormalization effects have been identified.

In contrast with previous studies, here we discuss 
the effect of renormalization on observables in terms of mass matrix.
The matrix of corrections can be written as:
\begin{equation}
\Delta M \approx -\epsilon_K M^0 - k 
\begin{pmatrix}2 \epsilon_e M_{ee}^0 & 
(\epsilon_e + \epsilon_\mu) M_{e\mu}^0 & 
(\epsilon_e + \epsilon_\tau) M_{e\tau}^0 
\cr
\dots & 2 \epsilon_\mu M_{\mu\mu}^0 & 
(\epsilon_\mu + \epsilon_\tau) M_{\mu\tau}^0 
\cr
\dots & \dots & 2 \epsilon_\tau M_{\tau\tau}^0
\end{pmatrix},
\label{moma}
\end{equation}
where $M^0$ is the matrix at the scale $m_0$ and $\epsilon_{K,e,\mu}$
can be obtained from the expression~(\ref{epsi}), by substituting
$y_\tau^2$ with $K,y_e^2,y_\mu^2$ respectively.

The effect on a given observable depends on how strong is the
influence (``imprint'') of this observable on the structure of the
matrix.  When two eigenstates are almost degenerate in mass, the
matrix structure depends very weakly on their mass squared difference
$\Delta m^2_{ij}$ and on the mixing angle
$\theta_{ij}$~\cite{MAN,MAI}.  In other words, $\Delta m^2_{ij}$ and
$\theta_{ij}$ are not ``imprinted'' in the matrix structure.  In this
case the small radiative corrections~(\ref{moma}) can strongly modify
these observables.  Let us give a rough estimation: $\theta_{ij}$ and
$\Delta m^2_{ij}$ can receive large radiative corrections if the
absolute neutrino mass scale $m$ is large enough to satisfy $\Delta
m^2_{ij}\sim \epsilon_\tau~m^2$. Because of the smallness of
$\epsilon_\tau$, this condition requires quasi-degenerate neutrino
masses.

In section~\ref{appli}, we will use eq.~(\ref{moma}) to discuss the
effect of radiative corrections for some particular structures of
$M^0$.

\section{Renormalization of the mass matrix due to new particles}\label{newp}

It may happen that new physics exists at some intermediate scale $m_X$
in the range $m_Z - m_0$, which does not contribute to neutrino masses
at tree-level but leads to renormalization effects.

\subsection{Non-standard Yukawa interactions}\label{scalep}

Let us consider new fermions and scalar bosons, with mass $\sim m_X$.
We assume that new scalars have zero VEV's, otherwise they would
generate neutrino masses at the scale $m_X<m_0$.  We analyze radiative
corrections to the operator~(\ref{five}) induced by the couplings of
these scalars and fermions to the lepton doublets $L_\alpha$:

\begin{itemize}
\item An extra scalar doublet, $\phi'$, with the coupling 
\begin{equation}
Y_{\alpha\beta}^{\phi'} \overline{L}_\alpha \phi' l_{R\beta} +
\hbox{h.c.}\,,
\label{fip}
\end{equation}
contributes to the wavefunction renormalization (diagram analogue to
figure~\ref{stand}$a$).  There is no vertex diagram with $\phi'$ in
the loop (analogue to figure~\ref{stand}$b$), because only $\phi$
enters the operator~(\ref{five}).

\item An extra scalar singlet $\rho$ or triplet $\chi$ 
can couple with two lepton doublets:
\begin{eqnarray}
&&
Y_{\alpha\beta}^{\rho}\overline{L}_\alpha i\sigma_2 L^c_\beta\rho 
+ \hbox{h.c.}\,,
\label{veroq}\\
&&Y_{\alpha\beta}^{\chi}\overline{L}_\alpha i\sigma_2 \sigma^i
L^c_\beta\chi_i + \hbox{h.c.}\,.
\label{veroq2}
\end{eqnarray}
The renormalization of the operator~(\ref{five}) is induced by the two
diagrams in figure~\ref{vero}.  In the singlet case, the matrix $Y$ is
antisymmetric, because $\overline{L}_\alpha i\sigma_2 L^c_\beta =
-\overline{L}_\beta i\sigma_2 L^c_\alpha$ (in the triplet case, $Y$ is
symmetric).

\FIGURE[t]{\epsfig{file=vero.eps,width=.6\textwidth}%
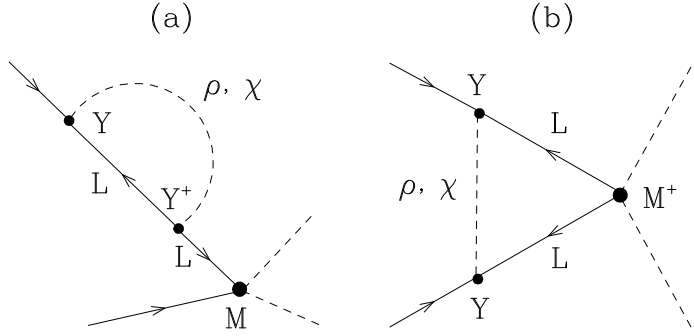
\caption{The diagrams generated by a new scalar singlet $\rho$ or
triplet $\chi$ (eqs.~(\ref{veroq}),~(\ref{veroq2})).\label{vero}}}

\item If a new right-handed doublet fermion $D_R$ exists at the scale
$m_X$, $\rho$ and $\chi$ can have the following interactions:
\begin{eqnarray}
&&Y_{\alpha}^{D\rho}\overline{L}_{\alpha}D_R\rho + \hbox{h.c.}\,,
\label{slc}\\
&&Y_{\alpha}^{D\chi}\overline{L}_{\alpha}\sigma^i D_R\chi_i + \hbox{h.c.}~\,.
\label{slc2}
\end{eqnarray}
In this case there are no vertex corrections, because $D_R$ and $\rho$
($\chi$) do not enter in the operator~(\ref{five}).  Only the diagram
shown in figure~\ref{right} contributes.

\FIGURE[t]{\centerline{\epsfig{file=right.eps,width=.4\textwidth}}%
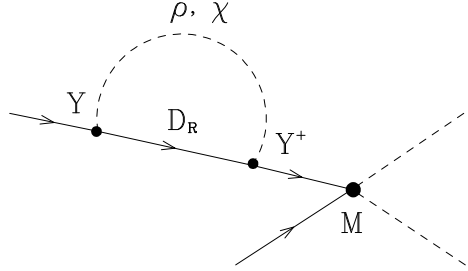
\caption{The diagram generated by a new scalar singlet $\rho$ 
(triplet $\chi$) and a new fermion doublet $D_R$ 
(eqs.~(\ref{slc}),~(\ref{slc2})).\label{right}}}

\item
A right-handed fermion singlet $S_R$, with the Yukawa interaction
\begin{equation}
Y_{\alpha}^S\overline{L}_{\alpha}\phi S_R + \hbox{h.c.}\,.
\label{psir}
\end{equation}
contributes both to wavefunction and vertex renormalization, as shown
in figure~\ref{newSL}. The scalar doublet in the
diagram~\ref{newSL}(a) can be the SM one, $\phi$, or some new
doublet~$\phi'$.

\FIGURE[t]{\epsfig{file=newSL.eps,width=.6\textwidth}%
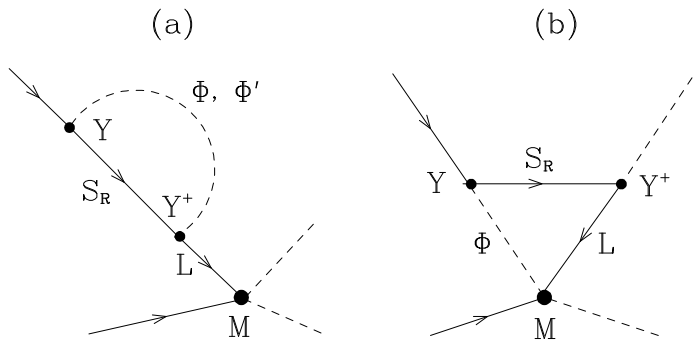
\caption{The diagrams generated by a new fermion singlet $S_R$
(eq.~(\ref{psir})).\label{newSL}}}

\end{itemize}
The new chiral fermions ($D_R$, $S_R$) can lead to anomalies, however
one can consider them as part of new vector-like fermions.  The vertex
diagrams in figures~\ref{vero}$b$ and~\ref{newSL}$b$ give no
contributions in the supersymmetric case, because of
non-renormalization theorem.

For any of the interactions in eqs.~(\ref{fip})--(\ref{psir}), 
the contribution to the $\beta$-function of neutrino mass matrix 
elements can be written as
\begin{equation}
16\pi^2(\beta_M^{Y})_{\alpha\beta}= k_Y^{(1)} \left[
M (Y^\dag Y) + (Y^\dag Y)^T M \right]_{\alpha\beta} + k_Y^{(2)} \left[
Y M^\dag Y^T \right]_{\alpha\beta}\,.
\label{sl2}
\end{equation}
where $Y$ are the new particle Yukawa couplings and the prefactors
$k_Y^{(i)}$ depend on the type of particles and interactions
considered.  The first term in square brackets of eq.~(\ref{sl2})
(analogue to those in eq.~(\ref{beta})) corresponds to all diagrams
discussed above (analogue to those in figure~\ref{stand}), except the
diagram in figure~\ref{vero}$b$.  This diagram generates the second
term in square brackets of eq.~(\ref{sl2}).  Therefore, $k_Y^{(2)}\ne
0$ only for the interactions in eqs.~(\ref{veroq}),~(\ref{veroq2}).

Let us emphasize the main differences in the running with respect 
to SM (MSSM):

\begin{itemize}

\item If the matrix $Y^\dag Y$ (or, in the case of
eqs.~(\ref{veroq}),~(\ref{veroq2}), the matrix $Y$) is not diagonal in
flavor basis, the RGE's for different matrix elements are coupled.  A
given matrix element receives corrections proportional to other matrix
elements; in this way small elements can be modified significantly.

\item The size of corrections depends on the size of Yukawa couplings
$Y$.  It can be much larger than in SM. In the perturbative regime,
$Y\lesssim 1$, the effect can be of the order of few percents, as in
MSSM.

\item The corrections can be further enhanced if several non-standard
multiplets are\linebreak present. E.g., one can introduce three generations of
new particles. In this case corrections can be as large as $\sim 10\%$
of the large matrix elements.

\item The size of corrections depends on $m_X$. Corrections are
suppressed for $m_X\sim m_0$.

\item In all diagrams but figure~\ref{vero}$b$, only one external
lepton leg is involved in flavor changing interactions. As a
consequence, a given element $M_{\alpha\beta}$ receives contributions
proportional to matrix elements in the rows (columns) $\alpha$ and
$\beta$ only.  In contrast, if the diagram in figure~\ref{vero}$b$ is
present, all matrix elements can contribute to the renormalization of
a given element $M_{\alpha\beta}$.

\end{itemize}
Since radiative effects are small, with a good approximation we can
consider only lowest order corrections to matrix elements. Using
eq.~(\ref{sl2}), we get:
\begin{equation}
\Delta M_{\alpha\beta}^Y \approx -\dfrac{\log(m_0/m_X)}{16\pi^2}
\left(k_Y^{(1)} \left[M(Y^\dag Y) + (Y^\dag Y)^TM
\right]_{\alpha\beta} +k_Y^{(2)} \left[Y M^\dag Y^T
\right]_{\alpha\beta} \right).
\label{deltac}
\end{equation}

\subsection{Non-universal $\U(1)$ gauge interaction}\label{u1x}

Let us consider the effect of extra heavy vector bosons $X_\mu$, which
have flavor non-universal interactions.  These bosons can be related
to the existence of horizontal (flavor) gauge symmetries~\cite{fn}.

We restrict ourselves to the case of an extra $\U(1)_X$ gauge group.
In the flavor basis, the interaction of lepton doublets with the new
gauge bosons $X_\mu$ has the form
\begin{equation}\
g_X Q_{\alpha\beta} \overline{L}_\alpha \gamma^\mu X_\mu L_\beta\,,
\label{gaco}
\end{equation}
where $g_X$ is the gauge coupling and $Q$ is the hermitian 
matrix of ``charges'', which can be non-diagonal in flavor basis.
The possible anomalies of the extra $\U(1)_X$ gauge group can be canceled
using the Green-Schwarz mechanism~\cite{gs}.

\FIGURE[t]{\epsfig{file=gauge.eps,width=.6\textwidth}%
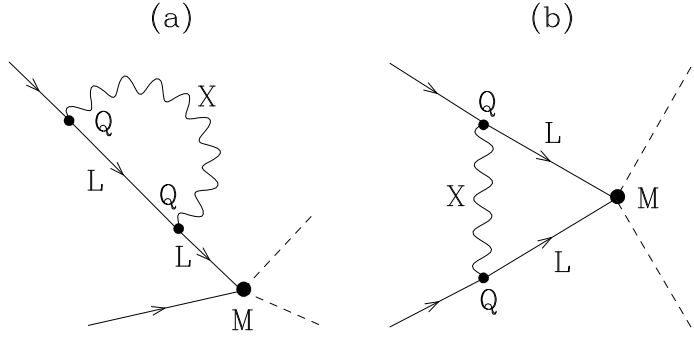
\caption{The diagrams generated by a new gauge boson $X_\mu$
(eq.~(\ref{gaco})).\label{gauge}}}

In figure~\ref{gauge} we show the two 1-loop gauge diagrams
that give non-universal 
radiative corrections to the operator~(\ref{five}).
Their contribution to the $\beta$-function of matrix elements takes 
the following form:
\begin{equation}
16\pi^2(\beta_M^{X})_{\alpha\beta}= g_X^2
k_X^{(a)} [M~Q^2 + (Q^2)^T M]_{\alpha\beta} + g_X^2 k_X^{(b)}
[Q^T M~Q]_{\alpha\beta}\,.
\label{bx2}
\end{equation}
The first and the second square brackets of eq.~(\ref{bx2}) are the
contribution of the diagrams in figure~\ref{gauge}$a$
and~\ref{gauge}$b$, respectively.  The $\beta$-function~(\ref{bx2})
has analogous features to the one in eq.~(\ref{sl2}): the RGE's of
different elements are coupled because of the off-diagonal entries in
the matrix $Q$; the corrections are proportional to the gauge coupling
$g_X^2$ and cannot be larger than few percents of the largest element
in $M$.  Like in figure~\ref{vero}$b$, in figure~\ref{gauge}$b$ both
lepton external legs enter in the loop.  As a consequence, a given
element $M_{\alpha\beta}$ can receive contributions proportional to
all matrix elements.  In lowest order, corrections can be written as:
\begin{equation}
\Delta M_{\alpha\beta}^X \approx -\dfrac{g_X^2\log(m_0/m_X)}{16\pi^2}
\left(k_X^{(a)} [M~Q^2 + (Q^2)^T M]_{\alpha\beta} +  k_X^{(b)}
[Q^T M~Q]_{\alpha\beta}\right).
\label{deltag}
\end{equation}

\section{Radiative generation of the sub-dominant  matrix elements}\label{appli}

The values of mass squared differences and mixing angles, measured in
neutrino oscillation experiments~\cite{exp}, are ($90\%$ C.L.):
\begin{eqnarray}
\begin{array}[b]{rclrcll}
\Delta m^2_{\rm sol}&\equiv&\Delta m^2_{12} =
\left( 7^{~ +10}_{~ -2} \right) \cdot 10^{-5} {\rm eV}^2\,,\qquad&
\tan^2 \theta_{12}&=&  0.42^{ + 0.2}_{ - 0.1}&
\hbox{(LMA MSW)}\,;\quad
\\ 
\Delta m^2_{\rm atm}&\equiv&\Delta m^2_{23}=
\left(2.5^{~ +1.4}_{~ -0.9} \right) \cdot 10^{-3} {\rm eV}^2\,,\qquad&
\tan \theta_{23} &=& 1 ^{~+ 0.35}_{~- 0.25}\,;
\qquad&
\sin\theta_{13} \lesssim 0.2\,.  
\end{array}   
\label{data}
\end{eqnarray}
Using these data, one finds that the neutrino mass matrix in flavor
basis can have a hierarchical structure, with some elements much
smaller than the others.  Small elements are suppressed by factors
$s_{13}$, $\Delta m^2_{\rm sol}/ \Delta m^2_{\rm atm}$ or, in the case
of degenerate spectrum ($m_1\approx m_2 \approx m_3$), by $\Delta
m^2_{\rm atm}/m_1^2$.  All possible hierarchical structures allowed by
the data have been identified in~\cite{MAI}.

Non-standard radiative effects can generate non-zero matrix elements
even if they are zero at the scale $m_0$.  These elements can receive
a radiative contribution up to $\sim 10\%$ of the largest matrix
element.  In what follows we will consider mass matrices with various
hierarchical structures and study the possibility to generate all
small elements of these matrices radiatively.  It has been found that
matrices with three or more exactly zero elements at the scale $m_Z$
cannot reproduce the experimental data~\cite{BGM}. We will show that
these matrices, realized at $m_0$, agree with phenomenology if
non-standard radiative corrections are taken into account.

Let us write the neutrino mass matrix with hierarchical structure as
$$
M=M_D+M_S\,,
$$
where $M_D$ is the matrix of dominant elements and $M_S$ is the matrix
of sub-dominant elements. In general, large matrix elements are the
sum of a dominant contribution from $M_D$ and a small correction from
$M_S$. Small matrix elements are contained in $M_S$ only.  We assume
that, at the high scale $m_0$,
\begin{equation}
M(m_0)=M_D(m_0)\,,\qquad  M_S(m_0)=0\,
\label{ms0}
\end{equation}
and, at the low scale $m_Z$,
\begin{equation}
M(m_Z)=M_D(m_0)+M_S(m_Z)\,,\qquad M_S(m_Z)=M_{\rm rad}\,,
\end{equation}
where $M_{\rm rad}$ is the matrix of radiative corrections.
In general, $M_{\rm rad}$ can be written as
$$
M_{\rm rad}=M_{SM(MSSM)}+M_{NS}\,,
$$
where $M_{NS}$ is the contribution of non-standard corrections.
In particular, we will analyze the corrections given by eq.~(\ref{deltac}),
assuming $k_Y^{(2)}=0$. In this case, $M_{\rm rad}$ can be written as
\begin{equation}
M_{\rm rad} = \lambda (X M_D + M_D X^T)\,,
\label{dema}
\end{equation}
where
\begin{equation}
X\equiv (Y^\dag Y)^T\,,\qquad
\lambda\equiv -\dfrac{k_Y^{(1)}}{16\pi^2}
\log\dfrac{m_0}{m_X}\,.
\label{xl}
\end{equation}

The non-standard (flavor changing) couplings $Y$ affect also the
evolution of the charged lepton Yukawa coupling matrix $Y_l$, inducing
non-zero off-diagonal entries.  As a consequence, flavor basis should
be redefined at the scale $m_Z$ through a rotation $U_l$ of charged
leptons.  Due to the strong hierarchy of charged lepton masses, the
mixing angles generated in the charged lepton sector are small, being
of the order of radiative corrections themselves:
$U_l=\mathbb{1}+U_{\rm rad}$.  Therefore, the neutrino mass matrix in
flavor basis is given by:
\begin{equation}
M^{fl}=U_l (M_D+M_{\rm rad}) U_l^T \approx 
M_D+M_{\rm rad}+U_{\rm rad}M_D+M_D U_{\rm rad}^T\,,
\label{charo}
\end{equation}
where we have neglected terms quadratic in radiative corrections.
While the neutrino masses at $m_Z$ can be extracted from $M=M_D+M_{\rm
rad}$, the mixing angles are influenced by $U_l$ rotation and all the
terms in eq.~(\ref{charo}) should be considered.

The term $(U_{\rm rad}M_D+M_D U_{\rm rad}^T)$ in eq.~(\ref{charo}),
describing the effect of charged leptons, has the same structure as
$M_{\rm rad}$ in eq.~(\ref{dema}).  Therefore, the effect of rotation
to flavor basis amounts to a redefinition:
\begin{equation}
X \rightarrow X + \frac{U_{\rm rad}}{\lambda}\,,
\label{shift}
\end{equation}
wheqre $X$ and $\lambda$ are defined in eq.~(\ref{xl}).
In the following, we will extract some results independent
from the form of $X$.

Let us comment on our assumption of exact zeros at $m_0$ (see
eq.~(\ref{ms0})).  Zero values of matrix elements can be a consequence
of certain symmetry at $m_0$.  The radiative contributions to these
elements are generated by interactions which break this symmetry.  In
general, the symmetry breaking leads to finite contributions already
at $m_0$.  On the other hand, if some particles producing radiative
corrections are light ($m_X\ll m_0$), the largest contribution to zero
elements is given by the leading logarithms and can be computed by RGE
methods in the context of the effective theory.  We are discussing
here these logarithmic contributions.

In what follows we will consider corrections to different dominant
structures $M_D$.

\subsection{Unit matrix}\label{iden}

Let us consider the matrix with dominant structure proportional 
to the unit matrix:
\begin{equation}
M_D = m\diag(1,1,1) \equiv m \mathbb{1}\,.
\label{ide}
\end{equation}
At $m_Z$, the approximate degeneracy of neutrino masses ($m_1\approx
m_2\approx m_3$) is broken by $\Delta m^2_{\rm atm}$ and $\Delta
m^2_{\rm sol}$.  Let us define
$$
\eta \equiv 1 - \dfrac{m_3}{m_2} \approx \pm \dfrac{\Delta m^2_{\rm atm}}
{2~m_1^2}\,,\qquad
\epsilon \equiv 1 - \dfrac{m_1}{m_2} \approx \dfrac{\Delta m^2_{\rm sol}}
{2m_1^2}\,,
$$
where the sign of $\eta$ is $+$ for inverted ordering ($m_3<m_2$) and
$-$ for normal ordering ($m_3>m_2$).  Taking $m_i \approx 0.3$eV (a
value allowed by all existing upper bounds~\cite{2beta,tritium,abso}),
one finds $\epsilon \ll \eta \lesssim 5\cdot10^{-3}$.  This means that
deviations from $M(m_Z)=m\mathbb{1}$ can be smaller than $1\%$.

For simplicity, at $m_Z$ we assume zero Majorana phases,
$\sin\theta_{13}=0$ and maximal atmospheric mixing
($\theta_{23}=\pi/4$). Then, the phenomenological mass matrix can be
written as~\cite{MAI}:
\begin{equation}
M^{ph}(m_Z) = m_2 \left(\begin{array}{ccc}
1-\epsilon c_{12}^2 & \epsilon c_{12}s_{12}/\sqrt{2} &  
- \epsilon c_{12}s_{12}/\sqrt{2}  
\\
\dots & 1-(\eta+\epsilon s_{12}^2)/2 & (-\eta+\epsilon s_{12}^2)/2 \\
\dots & \dots & 1-(\eta+\epsilon s_{12}^2)/2\\
\end{array}\right),
\label{mphe}
\end{equation}
where $c_{12}\equiv\cos\theta_{12}$ and $s_{12}\equiv\sin\theta_{12}$.

Let us consider, first, radiative corrections in the case of SM
(MSSM).  Substituting the matrix~(\ref{ide}) in eq.~(\ref{moma}), it
is evident that off-diagonal elements remain zero and, therefore, no
mixing is produced. Standard corrections lead to a split among
diagonal elements.  Using eq.~(\ref{moma}), one finds that mass
squared differences in the range required by phenomenology can be
obtained at $m_Z$ in the case of MSSM with large $\tan\beta$, however
one gets the following prediction:
$$ 
\dfrac{\Delta m^2_{\rm sol}}{\Delta m^2_{\rm atm}} \approx
\dfrac{\epsilon_\mu - \epsilon_e}{\epsilon_\tau - \epsilon_\mu} \approx
\dfrac{m_\mu^2}{m_\tau^2} \approx 3.5\cdot 10^{-3}\,.
$$
This ratio is about ten times smaller than the best fit value 
$\approx 2.5\cdot 10^{-2}$.

Let us consider now radiative corrections which originate from new
scalars and fermions (section~\ref{scalep}).  Substituting the
matrix~(\ref{ide}) in eq.~(\ref{dema}), we get corrections to matrix
elements of the following form:
\begin{equation}
M_{\rm rad} \approx 2~m~\lambda  \left(
\begin{array}{ccc}
X_{ee} & {\rm Re} X_{e\mu} & {\rm Re} X_{e\tau} \\
\dots & X_{\mu\mu} & {\rm Re} X_{\mu\tau} \\
\dots & \dots & X_{\tau\tau}\\
\end{array}\right),
\label{patt}
\end{equation}
where $X$ and $\lambda$ are defined in eq.~(\ref{xl}).  Since the
matrix $X$ is hermitian, the corrections to all matrix elements are
real, so that no CP violation is induced in this case.

The corrections to the different elements are related if a specific form
of the matrix $X$ is given. We consider 
an interaction like those in eqs.~(\ref{slc})--(\ref{psir}).  
Then, $X_{\alpha\beta}=Y_\alpha Y^*_\beta$ and, defining
\begin{equation}
\lambda_\alpha \equiv 
2 \lambda |Y_{\alpha}|^2\,,
\qquad c_{\alpha\beta} \equiv 
\cos(\arg{Y_{\alpha}}-\arg{Y_{\beta}})\,,
\qquad \alpha=e,\mu,\tau\,,
\label{lc}
\end{equation}
we get
\begin{equation}
M(m_Z)\equiv M_D+M_{\rm rad} = m \left(
\begin{array}{ccc}
1+\lambda_e & \sqrt{\lambda_e\lambda_\mu}c_{e\mu} &  
\sqrt{\lambda_e\lambda_\tau}c_{e\tau}  \\
\dots & 1+\lambda_\mu & \sqrt{\lambda_\mu\lambda_\tau}c_{\mu\tau} \\
\dots & \dots & 1+\lambda_\tau\\
\end{array}\right).
\label{would}
\end{equation}
The small parameters $\lambda_\alpha$ are real and have all the same
sign, which is opposite to the sign of $k_Y^{(1)}$ (see
eq.~(\ref{xl})).  The parameters $c_{\alpha\beta}$ can vary between
$-1$ and $1$, but only two of them are independent.

Before comparing the matrix~(\ref{would}) with the phenomenological
mass matrix~(\ref{mphe}), one needs to perform an additional rotation
$U_l$.  If $U_l$ is real, substituting $M_D\propto \mathbb{1}$ in
eq.~(\ref{charo}) we get
$$ 
M^{fl} = M_D + M_{\rm rad} + {\cal O}(U_{\rm rad}M_{\rm rad})\,. 
$$
In this case, we can neglect the rotation $U_l$.

\pagebreak[3]

Let us show that the matrix~(\ref{would}) can reproduce the
matrix~(\ref{mphe}).  For simplicity, we take $m=m_2$. Then, from the
condition $M_D+M_{\rm rad}=M^{ph}(M_Z)$, we find:
\begin{eqnarray}
\lambda_e &=& -\epsilon c_{12}^2\,,
\nonumber\\
\lambda_\mu&=&\lambda_\tau=-\dfrac 12 (\eta+\epsilon s_{12}^2)
\approx-\dfrac{\eta}{2}\,,
\nonumber\\
c_{\mu\tau}&=&\dfrac{\eta-\epsilon s_{12}^2}{\eta+\epsilon s_{12}^2}
\approx 1\,,
\nonumber\\
c_{e\mu}&=&-c_{e\tau}=\dfrac{\sqrt{\epsilon}~s_{12}}
{\sqrt{\eta+\epsilon s_{12}^2}} \approx \sqrt{\dfrac{\epsilon}{\eta}}~s_{12}\,.
\label{equa}
\end{eqnarray}
Since $\lambda_e$ is negative, also $\lambda_\mu$ and $\lambda_\tau$
are negative.  This corresponds to inverted mass spectrum ($\eta>0$).
To satisfy eq.~(\ref{equa}), one needs a positive $k_Y^{(1)}$.

If $m$ is not equal to $m_2$, other solutions are possible. It turns
out that the matrices~(\ref{mphe}) and~(\ref{would}) can be equal only
if $m$ is one of the three eigenvalues.  Requiring
$m=m_1\equiv(1-\epsilon)m_2$, one gets
\begin{equation}
\lambda_e \approx \epsilon s_{12}^2\,,
\qquad
\lambda_\mu=\lambda_\tau \approx -\dfrac{\eta}{2}\,,
\qquad
c_{\mu\tau} \approx 1\,,
\qquad
c_{e\mu}=-c_{e\tau} \approx \sqrt{-\dfrac{\epsilon}{\eta}}~c_{12}\,.
\end{equation}
This corresponds to normal mass spectrum.  Requiring
$m=m_3\equiv(1-\eta)m_2$, one gets
\begin{equation}
\lambda_e \approx \eta\,,
\qquad
\lambda_\mu=\lambda_\tau \approx \dfrac{\eta}{2}\,,
\qquad
c_{\mu\tau} = -1\,,
\qquad
c_{e\mu}=-c_{e\tau} \approx \dfrac{\epsilon c_{12}s_{12}}{|\eta|}\,.
\label{m=m3}
\end{equation}
This can correspond to both normal and inverted mass spectrum.

\subsection{Dominant $M_{ee}$ and $M_{\mu\tau}$}\label{eemt}

Let us consider the hierarchical matrix with dominant block given by
\begin{equation}
M_D = m \left(
\begin{array}{ccc}
1 & 0 & 0 \\
0 & 0 & -1 \\
0 & -1 & 0 \\
\end{array}
\right)\,.
\label{m02}
\end{equation}
For this matrix $\theta_{23}=\pi/4$, $\theta_{13}=0$ and the
eigenvalues equal $(m,m,-m)$.  Substituting the matrix~(\ref{m02}) in
eq.~(\ref{moma}), one sees that SM (MSSM) radiative corrections
generate a solar mass squared difference: $\Delta m^2_{\rm sol}\approx
2 m^2 \epsilon_\tau |k|$. However, atmospheric mass squared difference
and solar mixing angle remain zero.

We assume that the relative phases among eigenvalues at $m_Z$ are the
same as at $m_0$, that is $(1,1,-1)$ and also the values of
$\theta_{23}$ and $\theta_{13}$ remain $\pi/4$ and $0$,
respectively. Then, the phenomenological mass matrix is real and can
be written as~\cite{MAI}:
\begin{equation}
M^{ph}(m_Z) = m_2 \left(\begin{array}{ccc}
1-\epsilon c_{12}^2 & \epsilon c_{12}s_{12}/\sqrt{2} &  
- \epsilon c_{12}s_{12}/\sqrt{2}  \\
\dots & (\eta-\epsilon s_{12}^2)/2 & -1+(\eta+\epsilon s_{12}^2)/2 \\
\dots & \dots & (\eta-\epsilon s_{12}^2)/2\\
\end{array}\right),
\label{mphe2}
\end{equation}
where $\epsilon$, $\eta$, $c_{12}$ and $s_{12}$ are defined as in
section~\ref{iden}.

Let us study non-standard radiative corrections.  Substituting the
matrix~(\ref{m02}) in eq.~(\ref{dema}), we get
\begin{equation}
M(m_Z)\equiv M_D+M_{\rm rad}=m\left(\begin{array}{ccc}
1+\lambda_1 & \lambda_2 & -\lambda_2^*\\
\dots & \lambda_3 & -1+\lambda_4\\
\dots & \dots & \lambda_3^*\\
\end{array}\right),
\label{lax}
\end{equation}
where, using the definitions in eq.~(\ref{xl}), 
$$
\lambda_1 = 2 \lambda X_{ee}\,,\qquad
\lambda_2 = \lambda (X_{\mu e}-X_{e\tau})\,,\qquad
\lambda_3 = -2 \lambda X_{\mu\tau}\,,\qquad
\lambda_4 = - \lambda (X_{\mu\mu}+X_{\tau\tau})\,.
$$
To reproduce the phenomenological matrix~(\ref{mphe2}) with the
matrix~(\ref{lax}), one needs to take $\lambda_2$ and $\lambda_3$ real
($\lambda_1$ and $\lambda_4$ are real by definition, since $X$ is
hermitian).  Then, assuming for simplicity $m=m_2$, the equation
$M(m_Z)=M^{ph}(m_Z)$ leads to the following relations:
\begin{equation}
\tan2\theta_{12}=\dfrac{2\sqrt{2}\lambda_2}{\lambda_3-\lambda_1-\lambda_4}\,,
\qquad
\eta=\lambda_3+\lambda_4\,,
\qquad
\epsilon=\lambda_4-\lambda_1-\lambda_3\,.
\label{mm2}
\end{equation}

Let us consider the specific form
$X_{\alpha\beta}=Y_{\alpha}Y_{\beta}^*$, where $Y_\alpha$ are
non-standard Yukawa couplings.  In general, before a comparison with
eq.~(\ref{mphe2}), this form of $X$ should be modified as in
eq.~(\ref{shift}).  Assuming that the effect of this redefinition is
very small or can be reabsorbed in a redefinition of $Y_\alpha$, one
can express the equality $M(m_Z)=M^{ph}(m_Z)$ in terms of the
parameters defined in eq.~(\ref{lc}).  In particular, the equality of
matrices is realized for $m=m_3$ and the same values of parameters as
in eq.~(\ref{m=m3}).

Non-standard radiative corrections to the matrix~(\ref{m02})
has been discussed also in~\cite{ma}, in connection with a non-abelian
discrete symmetry that leads to the matrix~(\ref{m02}) at high energy.

\subsection{Dominant $M_{ee}$ and $\mu\tau$-block}\label{IH}

Let us assume that the
matrix with dominant block at the scale $m_0$ is given by
\begin{equation}
M_D = m \left(
\begin{array}{ccc}
1 & 0 & 0 \\
0 & 1/2 & -1/2 \\
0 & -1/2 & 1/2 \\
\end{array}
\right).
\label{m03}
\end{equation}
This matrix leads to $\theta_{23}=\pi/4$, $\theta_{13}=0$ and 
the eigenvalues equal $(m,m,0)$. It corresponds to inverted mass spectrum.
The SM (MSSM) radiative corrections  do not generate $1-2$ mixing because
the zero elements $M_{e\mu}$ and $M_{e\tau}$ are not modified by these 
corrections.

Let us consider the phenomenological mass matrix with the dominant
block structure~(\ref{m03}). We assume, for simplicity,
$m_3=\arg(m_1/m_2)=\theta_{13}=0$ and $\theta_{23}=\pi/4$. Then, the
matrix can be written as~\cite{MAI}:
\begin{equation}
M^{ph}(m_Z) = m_2 \left(
\begin{array}{ccc}
1-\epsilon c_{12}^2 & \epsilon c_{12}s_{12}/\sqrt{2} &  
- \epsilon c_{12}s_{12}/\sqrt{2}  \\
\dots & (1-\epsilon s_{12}^2)/2 & -(1-\epsilon s_{12}^2)/2 \\
\dots & \dots & (1-\epsilon s_{12}^2)/2\\
\end{array}\right),
\label{mphe3}
\end{equation}
where $m_2=\sqrt{\Delta m^2_{\rm atm}}\approx 0.05$eV and $\epsilon
\approx \Delta m^2_{\rm sol}/(2\Delta m^2_{\rm atm}) \approx 0.01$.

Let us study non-standard radiative corrections.  Substituting the
matrix~(\ref{m03}) in eq.~(\ref{dema}) and requiring
$M_{e\mu}=-M_{e\tau}$ and $M_{\mu\mu}=M_{\tau\tau}$ (necessary to
reproduce the matrix~(\ref{mphe3})), we get
\begin{equation}
M(m_Z)\equiv M_D+M_{\rm rad}=m\left(
\begin{array}{ccc}
1+\lambda_1 & \lambda_2 & -\lambda_2\\
\dots & 1/2+\lambda_3 & -1/2-\lambda_3\\
\dots & \dots & 1/2+\lambda_3\\
\end{array}
\right),
\label{lax2}
\end{equation}
where, using the definitions in eq.~(\ref{xl}), 
$$
\lambda_1 = 2 \lambda X_{ee}\,,\qquad
\lambda_2 = 2 \lambda {\rm Re}{X_{e\mu}}\,,\qquad
\lambda_3 = \lambda (X_{\mu\mu}-X_{\mu\tau})\,.
$$
Then, assuming for simplicity $m=m_2$, the equation $M(m_Z)=M^{ph}(m_Z)$  
leads to the following relations:
\begin{equation}
\lambda_2=\sqrt{\lambda_1\lambda_3}\,,\qquad
\tan^2\theta_{12}=\dfrac{2\lambda_3}{\lambda_1}\,,
\qquad
\epsilon=-\lambda_1-2\lambda_3\,.
\end{equation}

If the matrix $X$ has the specific form
$X_{\alpha\beta}=Y_{\alpha}Y_{\beta}^*$, one can express the equality
$M(m_Z)=M^{ph}(m_Z)$ in terms of the parameters defined in
eq.~(\ref{lc}).  For $m=m_2$, we get:
$$
\lambda_e=-\epsilon c_{12}^2\,,\qquad
\lambda_\mu=\lambda_\tau=-\frac{\epsilon s_{12}^2}{2}\,,\qquad
c_{\mu\tau}=-1\,,\qquad
c_{e\mu}=-c_{e\tau}=1\,.
$$

\subsection{Dominant $\mu\tau$-block}\label{NH}

Let us take, at the scale $m_0$, the matrix with dominant $\mu\tau$-block~\cite{domi}:
\begin{equation}
M_D=m
\left(
\begin{array}{ccc}
0 & 0 & 0\\
0 & 1 & 1\\
0 & 1 & 1\\
\end{array}
\right).
\label{domi}
\end{equation}
It corresponds to the case of normal hierarchical mass spectrum
($m_3\gg m_2\gg m_1$).

The mass matrix required by phenomenology can be written as an
expansion over the small parameters $\cos 2\theta_{23}$,
$\sin\theta_{13}$ and $r\equiv\sqrt{\Delta m^2_{\rm sol}/\Delta
m^2_{\rm atm}}$~\cite{MAN}:
\begin{eqnarray}
M^{ph}(m_Z) &\approx& \dfrac{m_3}{2}\left[
\left(
\begin{array}{ccc}
0 & 0 & 0\\
\dots & 1 & 1\\
\dots & \dots & 1\\
\end{array}
\right)+
\cos 2\theta_{23}\left(
\begin{array}{ccc}
0 & 0 & 0\\
\dots & -1 & 0\\
\dots & \dots & 1\\
\end{array}
\right)\right.+
\label{deco}\\&&
           \hphantom{ \dfrac{m_3}{2}\Biggl[}\!
\left. +
\sin\theta_{13}e^{i\delta}\left(
\begin{array}{ccc}
0 & \sqrt{2} & \sqrt{2}\\
\dots & 0 & 0\\
\dots & \dots & 0\\
\end{array}
\right)+ r e^{2i\sigma}\left(
\begin{array}{ccc}
2s_{12}^2 & \sqrt{2}s_{12}c_{12} & -\sqrt{2}s_{12}c_{12}\\
\dots & c_{12}^2 & -c_{12}^2\\
\dots & \dots & c_{12}^2\\
\end{array}
\right)
\right],
\nonumber
\end{eqnarray}
where $\delta$ is the CP violating Dirac phase and $\sigma$ is the CP
violating relative Majorana phase between $m_2$ and $m_3$.

Let us discuss, first, the effect of SM (MSSM) radiative corrections
to the matrix~(\ref{domi}).  Using eq.~(\ref{moma}), one sees that
$e$-row elements remain zero along the RGE running. It is a
consequence of eq.~(\ref{propto}).  In contrast, SM (MSSM) corrections
to $\mu\tau$-block elements are present.  These corrections induce a
small deviation from $\theta_{23}=\pi/4$, which can be easily computed
using eq.~(\ref{moma}):
$$
\cos2\theta_{23}=(\epsilon_\mu-\epsilon_\tau)k
+{\cal O}(\epsilon^2)\,.
$$
The first neutrino remains massless and unmixed.  Also $m_2$ remains
zero because the determinant of the $\mu\tau$-block is zero even after
the inclusion of radiative corrections.  Therefore, the solar mass
difference and mixing angle are not generated radiatively.

Let us consider, now, the effect of non-standard radiative corrections. 
Using the dominant matrix~(\ref{domi}), the matrix of corrections, 
given by eq.~(\ref{dema}), is:
\begin{equation}
M_{\rm rad}=m\left(\begin{array}{ccc}
0 & \lambda_1 & \lambda_1\\
\dots & 2\lambda_2 & \lambda_2+\lambda_3\\
\dots & \dots & 2\lambda_3\\
\end{array}\right).
\label{lambda}
\end{equation}
where, using the definitions in eq.~(\ref{xl}), 
$$
\lambda_1 = \lambda (X_{e\mu}+X_{e\tau})\,,\qquad
\lambda_2 = \lambda (X_{\mu\mu}+X_{\mu\tau})\,,\qquad
\lambda_3 = \lambda (X_{\tau\mu}+X_{\tau\tau})\,.
$$
The eq.~(\ref{dema}) implies that a given element $M_{\alpha\beta}$
receives corrections proportional only to elements in the $\alpha$ and
$\beta$ rows of $M_D$.  Since the first row and column in
eq.~(\ref{domi}) are zero, the element $M_{ee}$ remains zero.
Corrections to $M_{e\mu}$ and $M_{e\tau}$ are equal.  The matrix
$M=M_D+M_{\rm rad}$ has a zero eigenvalue, so that strong normal
hierarchy is preserved by radiative corrections.

Comparing the matrix of radiative corrections~(\ref{lambda}) with the
phenomenological matrix~(\ref{deco}), we find that the atmospheric
angle gets a deviation from $\pi/4$ and the angle $\theta_{13}$
becomes non-zero:
\begin{equation}
\cos 2\theta_{23}\approx \lambda_3-\lambda_2\,,\qquad
\sin\theta_{13} \approx \lambda_1 \frac{e^{-i\delta}}{\sqrt{2}}\,. 
\label{t123}
\end{equation}
A problem appears with the generation of solar parameters,
because the structure of the term
proportional to $r$ in eq.~(\ref{deco}) 
(let us call it $M^{\rm sol}$) 
differs from the structure~(\ref{lambda}) of radiative corrections:
$$
M^{\rm sol}_{ee}\ne 0\,,\qquad
M^{\rm sol}_{e\mu}=-M^{\rm sol}_{e\tau}\,,\qquad
M^{\rm sol}_{\mu\tau}=-M^{\rm sol}_{\mu\mu}=-M^{\rm sol}_{\tau\tau}\,,
$$
whereas 
$$
M^{\rm rad}_{ee}= 0\,,\qquad
M^{\rm rad}_{e\mu}=M^{\rm rad}_{e\tau}\,,\qquad
M^{\rm rad}_{\mu\tau}=\dfrac 12 (M^{\rm rad}_{\mu\mu}+M^{\rm rad}_{\tau\tau})\,.
$$
In fact, computing the eigenvalues of $M=M_D+M_{\rm rad}$, one finds
that $r$ is of order $\lambda_i^2$:
\begin{equation}
r\equiv \dfrac{m_2}{m_3} \equiv 
\sqrt{\dfrac{\Delta m^2_{\rm sol}}{\Delta m^2_{\rm atm}}}
\approx \dfrac{\lambda_1^2+(\lambda_2-\lambda_3)^2/2}{2}\,.
\label{rrr}
\end{equation}
To satisfy the lower bound ($99\%$ C.L.) on $r$, which is $\sim 0.1$
(LMA), one needs $\lambda_i \gtrsim 0.2 \div 0.3$. In other words, LMA
solar mass difference can be generated only if radiative corrections
are $\sim 10$ times larger than in the MSSM with large $\tan\beta$.

The equalities $M_{ee}=0$ and $M_{e\mu}=M_{e\tau}(=m\lambda_1)$ in
eq.~(\ref{lambda}) lead to the following general relations between
observables~\cite{MAN}:
\begin{equation}
c_{12}^2= \frac{1+r}{2r}(1-\sin 2\theta_{23})\,,\qquad
\tan\theta_{13}= s_{12}\sqrt{r}\,,\qquad
\sin\delta= 0\,,\qquad
\sigma= \frac{\pi}{2}\,.
\label{pred}
\end{equation}
The first equality in~(\ref{pred}) shows that the LMA solar mixing
angle can be obtained only if atmospheric mixing deviates
significantly from maximal value and if $r$ is substantially smaller
than the present best fit value. The equality can be satisfied taking
$99\%$ C.L.  allowed intervals for $\theta_{12}$, $\theta_{23}$ and
$r$.  The other equalities in~(\ref{pred}) show that the radiative
corrections~(\ref{lambda}) give $s_{13}$ close to the present upper
bound $\sim 0.2$ and lead to CP conservation in oscillations and to
opposite CP parity between $m_2$ and $m_3$.

We have shown that, using the form~(\ref{lambda}) of corrections, the
predictions for $\Delta m^2_{\rm sol}$ and $\theta_{12}$ are too small
with respect to phenomenological values.  However, the corrections
related to the second term in square brackets of
eqs.~(\ref{deltac}),~(\ref{deltag}) have qualitatively different
features and can give better predictions for the solar parameters.  In
particular, a non-zero $M_{ee}$ can be generated by these corrections.

For example, let us consider the interaction in eq.~(\ref{veroq}).
The $\beta$-function coefficients are, in this case, $k_Y^{(1)}=1/2$
and $k_Y^{(2)}=1$.  The matrix of couplings $Y$ is antisymmetric,
therefore the matrix of corrections $\Delta M^Y$ (see
eq.~(\ref{deltac})) depends on three independent couplings only:
$Y_{e\mu}$, $Y_{e\tau}$ and $Y_{\mu\tau}$.  We want to compare
$M(m_Z)\equiv M_D+\Delta M^Y$ with the phenomenological mass
matrix~(\ref{deco}) (we neglect, for simplicity, the charged lepton
rotation $U_l$).  The equality of the two matrices can be realized for
$\sin\theta_{13}= \cos 2\theta_{23}=0$ and
\begin{equation}
m=\dfrac{m_3}{2}+\dfrac{m_2}{2}\,,\qquad
\lambda Y^2_{e\mu} = \lambda Y^2_{e\tau} = \dfrac{r s_{12}^2e^{2i\sigma}}
{4(1-r)}\,,\qquad
 \lambda Y^2_{\mu\tau} = \dfrac{r c_{12}^2 e^{2i\sigma}}
{2(1-r)}\,,
\label{goodpred}
\end{equation}
where $\lambda$ is defined in eq.~(\ref{xl}).

Notice that, while the form~(\ref{lambda}) of corrections predicts $r$
of order $\lambda_i^2$, here $r$ is of order $\lambda_i$.  Therefore,
$\sim 10\%$ radiative corrections are enough to generate radiatively
$\Delta m^2_{\rm sol}$.  Such corrections can be produced if three
generations of new particles (e.g., three scalar singlets) are
introduced.  As follows from eq.~(\ref{goodpred}), also the solar
mixing angle $\theta_{12}$ can be generated radiatively (in the LMA
allowed range).

\section{Discussion and conclusions}

The structure of the Majorana neutrino mass matrix can be
reconstructed, using experimental data, at the electroweak scale
$m_Z$.  This structure changes with RGE running to the high energy
scale $m_0$.  We have analyzed the features of the running between the
two scales.

The SM (MSSM) radiative corrections do not modify the matrix
structure.  In flavor basis, the value of each matrix element at $m_Z$
is proportional to the one at $m_0$. Moreover, the corrections to this
value cannot be larger than few percents. Therefore, both the dominant
structure of the mass matrix and the small matrix elements are not
modified significantly between $m_0$ and $m_Z$. Zero elements remain
zero.

At the same time SM and MSSM corrections can change significantly
observables: corrections can enhance or suppress mixing, modify
strongly $\Delta m^2$ or even generate mass split.  Substantial change
of observables occurs for the quasi-degenerate spectrum, with common
scale of neutrino mass $m \gtrsim 0.1$\,eV.

We have studied the radiative effects induced by new particles and
interactions at a scale $m_X$, with $m_Z < m_X < m_0$.  These
non-standard (flavor changing) corrections lead to coupled RGE's of
different matrix elements. As a consequence, small tree-level elements
get corrections proportional to the large matrix elements.  We have
considered non-standard corrections induced by new scalar bosons, new
fermions and new gauge bosons.
 
In all cases, the dominant structure of the mass matrix remains the
same between $m_Z$ and $m_0$. Therefore, if the matrix structure at
$m_0$ is determined by some symmetry, this symmetric structure can be
identified from the experimental data at $m_Z$.  However, the values
of small elements can be strongly modified.  We show that small
elements of hierarchical matrices can be zero at the scale $m_0$ and
receive non-zero contributions from radiative corrections.  At the
high mass scale, only the dominant block elements can be non-zero.

In the case of exactly degenerate neutrino masses, small ($\sim (0.1
\div 1) \%$) corrections to zero elements can generate large mixing
angles and mass squared differences in the range required by
phenomenology, both for solar and atmospheric neutrinos.  We have
shown, in particular, that the unit matrix can be the exact form of
the neutrino mass matrix at $m_0$.

In the case of inverted hierarchy, the structure with zero $e\mu$ and
$e\tau$ elements at $m_0$ can lead to correct predictions for low
energy solar parameters, if $1\%$ non-standard corrections are
present.
 
In the case of normal hierarchical neutrino masses, we have studied
the matrix with dominant $\mu\tau$-block.  Solar mass squared
difference and mixing angle can get large renormalization effects. To
generate a mass difference in the LMA region, one needs $10\%$
radiative corrections.

\acknowledgments

We would like to thank G.~Senjanovic for useful discussions. M.~F.
thanks also S.~Bertolini and D.~Perini for many comments and clarifications.
The work of M.~F. is supported in part by the Italian MIUR under 
the program ``Fenomenologia delle Interazioni Fondamentali''.

\end{document}